\documentclass[12pt]{article}
\usepackage[letterpaper,left=1.2in,right=1.2in,top=1.2in,bottom=1.2in]{geometry} 
\usepackage[latin1]{inputenc} 
\usepackage{setspace} 
\usepackage{fancyhdr} 
\usepackage{amsmath} 
\usepackage{amssymb} 
\usepackage{amsfonts} 
\usepackage{dsfont} 
\usepackage{amsthm} 
\usepackage{graphicx} 
\usepackage{tikz} 
\usepackage{color} 
\definecolor{goetheblue}{RGB}{0,102,255} 
\usepackage{listings} 
\usepackage{eurosym} 
\usepackage{url} 
\usepackage{enumerate} 
\usepackage{multirow} 
\usepackage{verbatim} 
\usepackage{mathrsfs}
\usepackage{booktabs}
\usepackage{multirow}
\usepackage{adjustbox}		
\usepackage[round]{natbib}
\usepackage{url}
\input{ee.sty}        
\usepackage{bm}       

\usepackage{amsfonts}
\usepackage{amsbsy}
\usepackage{rotating}
\usepackage[flushleft]{threeparttable}
\usepackage[margin=10pt,font=small,labelfont=bf,labelsep=period]{caption}
\usepackage{arydshln}

\allowdisplaybreaks

\newcommand{\ba}{\begin{array}}
	\newcommand{\ea}{\end{array}}

\newcommand{\be}{\begin{equation*}}
\newcommand{\ee}{\end{equation*}}
\newcommand{\bea}{\begin{eqnarray*}}
	\newcommand{\eea}{\end{eqnarray*}}

\newcommand{\E}{\mbox{E}}
\newcommand{\p}{\mbox{P}}

\newtheorem{theorem}{Theorem}

\newtheorem{remark}[theorem]{Remark}

\renewcommand{\arraystretch}{1.5} 

\begin{document}
\title{When More Is Less:\\ Pitfalls of significance testing }\vspace{5mm} 
\vspace{5mm} 
\author{Uwe Hassler\footnote{I thank Tanja Zahn for coding and performing the computer experiments. I am grateful to J\"org Breitung, Mehdi Hosseinkouchack,  Jan Reitz, Jan-Lukas Wermuth and Michael Wolf for many helpful comments.} \\ %
\vspace{5mm} \emph{Goethe University Frankfurt}\thanks{Statistics and Econometric Methods, RuW Building, 60629 Frankfurt, Germany. Email:
	hassler@wiwi.uni-frankfurt.de}}
\date{ \today } 
\maketitle		

 \begin{abstract}
The controversy about statistical significance vs. scientific relevance is more than 100 years old. But still nowadays null hypothesis  significance testing is considered as gold standard in many empirical fields from economics and social sciences over psychology to medicine, and small $p$-values are often the key to publish in journals of high scientific reputation. I highlight,  illustrate and discuss  potential pitfalls of statistical significance testing on three occasions.


 
 	\noindent
 	\newline
 		\textbf{Keywords:} $p$-values; data snooping;  agnotology.
 	
 \end{abstract}

\section{Introduction} 

Public opinion expects scientific results to be correct and  relevant. Hence, statistically significant findings sound tempting and suggestive, although they are not necessarily  meaningful.  \citet{Boring1919} observed an antagonism ``Mathematical vs.~scientific significance'' and stated an ``apparent inconsistency between scientific intuition and mathematical result''. One hundred years later,  two prominent papers reconfirmed such early critique and expressed far more vehement objections. First, \citet{AmrheinGreenlandMcShane2019} gathered in \emph{Nature} more than 800 scientists behind their ``Rise up Against Statistical Significance''. They ``are not calling for a ban on $P$ values'', but ``are calling for a stop to the use of $P$ values
in the conventional, dichotomous way -- to
decide whether a result refutes or supports a
scientific hypothesis''. Second, \emph{The American Statistician} published a supplementary issue (Volume 73, 2019) on \emph{Statistical Inference in the 21st Century: A World Beyond} $p < 0.05$. It comprises more than 40 papers including the editorial expressing a more radical view (\citet[p.~2]{Wasserstein2019}): ``We conclude, based on our review of the articles in this special
issue and the broader literature, that it is time to stop using
the term `statistically significant' entirely. Nor should variants
such as `significantly different', `$p < 0.05$', and `nonsignificant'
survive, whether expressed in words, by asterisks in a table, or
in some other way.''  What is the reason behind the demanded rise against or even ban of statistical significance?

There is a rich tradition of critical comments and sceptical views on null hypothesis  significance testing, see e.\,g., \citet{Berkson1938}, \citet{Rozeboom1960}, \citet{Bakan1966} and \citet{Meehl1978}. Hence, \citet{Cohen1994} wrote in his paper with the polemic title \emph{The Earth Is Round} ($p < .05$): ``After 4 decades of severe criticism, the ritual of null hypothesis significance testing -- mechanical dichotomous
decisions around a sacred .05 criterion -- still persists.'' Consequently, after the turn of the century the criticism has become even more pronounced: \citet{Gigerenzer2004}, \citet{ioannidis2005} and \citet{ZiliakMcCloskey2008} headlined \emph{Mindless statistics},  \emph{Why most published research findings are false} and \emph{The cult of statistical significance}, respectively.

Given the vast literature on  statistical significance, it is not surprising that I have little to add that is new. Instead, I address the  reader who is not an expert in statistics and wish to point out three common pitfalls when statistical tests are applied in practice: Although data are not manipulated and true results are reported (``facts''), the conclusions are faulty (``fake''), because the methods are not applied in a correct manner. My examples quantify this effect of malfunctioning.

The next section briefly reviews the very nature of the ``ritual of significance testing''. Section\,3 discusses 3 examples that are representative of potential fallacies lurking in applied work. Conclusions are offered in the final section. The paper ends with technical remarks that guarantee formal correctness of some of the arguments but are not essential for the understanding of the main message.


\section{Proof by contra(pre)diction}

The concept of computing a $p$-value as criterion to judge whether some observations can be considered as generated by random or not has been formalized by \citet{Pearson1900}, although he did not use the term ``significant'' yet. The earliest use of ``statistical significance'' I am aware of (admittedly not being a historian) is by \citet[p.~315]{Boring1916}. It was the highly influential book by \citet{Fisher1925} that firmly anchored ``tests of significance'' and $p$-values in much of the empirical research over the last century, where $H_0$ signifies a null hypothesis to be tested. The index is motivated by (agricultural) treatment studies dominating in \citet{Fisher1925}: Under the null hypothesis there is a zero effect. What one takes for granted is a collection of data, from which the value of a test statistic is computed, say $t$; the larger $t$, the less likely is the occurrence of this value if $H_0$ is true. Let $T$ denote the test statistic. The $p$-value is defined as the probability to observe a value as large as $t$ or even larger, assuming that $H_0$ is true:
\[
P := \p ( T \geq t | H_0) \, .
\]
Hence, $P$ is the conditional probability to observe the value $t$ of the test statistic, or an even stronger violation of $H_0$, given the null hypothesis is true. By indirect conclusion, the trust in the truth of $H_0$ is shattered for small $P$ or confirmed for large $P$ by the data summarized in $t$: $P$ is used to formulate confidence in the assumed null hypothesis.

In mathematics, the method of proof by contradiction works as follows: $C$ is certain, taken for granted; $A$ is assumed; $B$ follows from $A$ and contradicts $C$; hence the assumption $A$ must be wrong. Statistical significance tests work along similar lines: $C$ is certain, namely the data that we observe; $A$ is the null hypothesis; assuming $A$ to hold, one computes the probability $P$ with which $C$ is predicted from $A$; the smaller $P$ is, the less confidence we have in the $A$. Formally, if $P$ is ``very small'' one rejects the null hypothesis $A$, knowing of course that such a decision may be wrong. But the probability of a wrongful rejection is controlled by the probability $P$. 

To the best of my knowledge, the analogy between hypothesis testing and proof by contradiction has first been stressed by \citet{ReevesBrewer1980}. \citet[p.~75]{FalkGreenbaum1995} criticized this kind of statistical inference to be a ``flawed logical structure'' since ``the conclusion that, given a significant result, $H_0$ becomes improbable is not generally true''. Notice, in a ``classical'' (or: frequentist) view, the null hypothesis is a state of the world, which is either true or not. Such a view allows for stronger or weaker belief in $H_0$, but it does not model the degree of confidence by means of probabilities. On the contrary, a Bayesian take is to model the degree of belief or disbelief as prior probability, which is beyond the scope of our note.

\section{Three fallacies}

Data mining or data snooping techniques are widespread in empirical research. While such tools are perfectly legitimate  for some purposes, they are prone to invalidate statistical inference. Care and caution are hence advisable when it comes to significance testing after preliminary data analysis. Three examples show  that incorrect application of statistical tests may produce misleading results.

\subsection{$p$-hacking}

Before turning to a real case, the paper by \citet{Bem2011}, I want to consider an artificial example.

{\sc Example~1} \emph{Consider the following (computer) experiment. We draw a sample of size $n=10^5$ from (pseudo) random numbers $1, 2, \ldots, 100$ and count how often repdigits (numbers with repeated digits) occur. There are 9 repdigits between 1 and 100, namely $E = \{11, 22, \ldots, 99\}$. Hence, one expects the event $E$ to occur with probability $\p(E) = \frac{9}{100} = 0.09$ when random sampling. The observed  frequency in the experiment, however, is considerably smaller: 8,804 cases out of $n=100,000$, i.e., the observed relative frequency is $\widehat{\p} (E) = 0.08804$, where $\widehat{\p} (\cdot) $ is short for the empirical relative frequency. What do I mean by ``considerably smaller''? Is the difference between $0.09$ and $0.088$ simply due to sampling variability and caused by chance? Or is this difference ``systematic''? In fact, it is ``statistically significant'', or more precisely  ``significant at the 5\,\% level'' in that a test statistic testing for the null hypothesis $\p(E) =0.09$ results in a $p$-value smaller than 5\,\%, see Remark~\ref{rem_1} at the end of the paper.}

There seems to be no good reason why  repdigits should occur less often in random samples than theoretically expected or predicted. So, what is going wrong in Example~1?  Most statisticians would say that a sample size of $n=10^5$ is not too small. The experiment was carried out with the open source software R, and searching the internet you will not find comments hinting at evidence  that the (pseudo) random number generator is defective; see also Remark~\ref{rem_2}. If nothing went wrong, should I submit my finding that repdigits occur significantly (at level 5\,\%) less often than theoretically predicted to a scientific journal? Is it possible to publish empirical findings hardly anyone has trust in a priori? Isn't it the very nature of science to dump an a priori hypothesis if data contradict it?

The last two questions have been discussed in connection with the (in)famous study by \citet{Bem2011} on extrasensory perception (ESP). 100 participants of an experiment had to guess, or rather predict, whether a picture would show up on the left or on the right of their screen in front of them. The position was determined randomly by computer with equal chances (50\,\%);  the pictures were of erotic or nonerotic content. \citet[p. 409]{Bem2011} clarifies that ``neither the picture itself nor its left/right position was determined until after the participant recorded his or her guess, making
the procedure a test of detecting a future event (i.e., a test of
precognition)''. He summarizes ``Across all 100 sessions, participants correctly identified the future position of the erotic pictures significantly more frequently than the 50\,\% hit rate expected by chance: 53.1\,\%'', resulting in (one-tailed) significance of 1\,\%. Further, ``In contrast, their hit rate on the nonerotic pictures did not differ significantly from chance: 49.8\,\%'' with a $p$-value larger than 0.5. Of course, I cannot tell how many  tests were executed (without rejection?) before the significant alternative of   erotic pictures showed up and the study got published in one of the leading journals of the American Psychological Association -- and it seems that Daryl Bem cannot tell either. He is quoted by \citet{Engber2017} as follows: ``I would start one [experiment], and if it just wasn't going anywhere, I would abandon it and restart it with changes'', and ``I didn't keep very close track of which ones I had discarded and which ones I hadn't''.  Note this is not a case of scientific fraud -- but maybe sloppiness. \citet{Engber2017} quotes Bem as  ``I think probably some of the criticism could well be valid. I was never dishonest, but on the other hand, the critics were correct.''

I now offer a way out of our discomforting empirical evidence given in Example~1. Similarly to Bem in \citet{Engber2017} I now admit that I actually  tested 20 nonsensical hypotheses at the 5\,\% level building on the events
\begin{eqnarray*}
E_1 &=& \{2, 4, 8, 16, 32, 64\}, \mbox{ i.\,e., powers of 2 with probability } \p(E_1) = \frac{6}{100}, \\
E_2 &=& \{3, 9, 27, 81\},  \mbox{ i.\,e., powers of 3 with } \p(E_2) = \frac{4}{100},\\
E_3 &=& \{2,3,5,8,13,21,34,55,89\},  \mbox{ i.\,e., Fibonacci numbers with } \p(E_3) = \frac{9}{100},\\
& \vdots&\\
E_{20} &=& \{2,3,5,7,11,13, \ldots , 83, 89, 97\},  \mbox{ i.\,e., prime numbers with } \p(E_{20}) = \frac{25}{100},
\end{eqnarray*}
including $E = \{11, 22, \ldots, 99\}$. In each case I confronted the theoretical null hypothesis $\p(E_j)$ with the empirical sample pendant $\widehat{\p} (E_j)$, and only the case of repdigits was significant (at 5\,\%). If you perform 20 tests at significance level 5\,\% and if all null hypotheses are true, then you must expect one test out of 20 tests to be erroneously significant at the 5\,\% level. This is part of the logic of ``conclusion by contraprediction'' of significance tests. Hence, in Example~1 I did not manipulate the data or the statistics but reported the empirical facts. I only concealed that a lot of tests were run before finding and reporting a significant result, which for this reason is actually fake. Such a research practice is related to so-called  $p$-hacking, see  \citet{Simonsohnetal2014} and \citet{Simmonsetal2011}. One continues collecting data or carrying out experiments until a sufficiently small $p$-value shows up, thus promising some significant result, which is prone to ``false positive'' findings. A related strategy is  HARKing (Hypothesizing After the Results are Known, see \citet{Kerr1998}) or SHARKing (secretly HARKing, see \citet{HollenbeckWright2017}): With data mining techniques, empirical or experimental data are screened until a striking feature shows up; having detected the feature one tests whether it is significant. This seems to be a widespread research practice: \citet{Johnetal2012}  carried out a survey among academic psychologists; they observed that one third of them admitted  cases of ``reporting an unexpected
finding as having been
predicted from the start''.


\subsection{Maximizing evidence subject to snooping}

The next example continues with a stylized case how flawed statistical significance may be produced when performing several tests on the same data and reporting only the most significant one.

{\sc Example 2} \emph{Assume a researcher observes in a pandemic situation two univariate time series $\left\{ x_t \right\}$ and $\left\{ y_t \right\}$ for $t=1, \ldots, n$ with $k-1$ additional starting values for $x_t$. He or she suspects that past or contemporaneous values $x_t$, $x_{t-1}$, ..., $x_{t-k+1}$  might help to explain $y_t$ such that (past) values of $x_t$ allow to forecast $y_t$. Suppose that forecasts of turning points in $y_t$ are of major relevance when trying to steer the pandemic. Hence, a leading indicator for the future development of $y$ is very desirable. If $x$ is such an indicator, one sometimes says that $x$ is (Granger) causal for $y$ after the seminal paper by \citet{Granger1969}. To establish forecastability or causality in this sense, we run a least squares (LS) regression:
\[
y_t = \widehat \beta_0 + \widehat \beta_1 x_t + \widehat \beta_2 x_{t-1} + \ldots + \widehat \beta_k x_{t-k+1} + \widehat u_t \, , \quad t= 1, \ldots, n \, .
\]
Consider the null hypotheses that $x_{t-h+1}$ does not help to predict $y_t$ ($h=1, \ldots, k$), i.\,e., $H_0^{(h)}$: $\beta_h = 0$. 
Let $T_h$ denote the standard $t$ ratio. Assuming normally distributed data it holds under the null that $T_h$ follows a $t$ distribution with $n-k-1$ degrees of freedom, $t(n-k-1)$. Under the maintained assumptions a valid two-tailed test at significance level $\alpha$ amounts to a rejection of $H_0^{(h)}$ if $|T_h | > t_{1-\alpha/2} (n-k-1)$. The null hypothesis of no Granger causality (given a prespecified value of $k$) is
\begin{equation*} \label{H0}
H_0 : \ \beta_1 = \ldots = \beta_k= 0 \ \mbox{ or } H_0^{(1)} \cap \ldots \cap H_0^{(k)}. 
\end{equation*}
If any $H_0^{(h)}$ is false, then the intersection $H_0$ does not hold, too. Driven by the strong wish or need to forecast the target variable $y$ the researcher maximizes the evidence in favour of causality by computing
\[
T_{max} := \max \left\{|T_1 |, \ldots, |T_k | \right\} ,
\]
which singles out one individual (namely the most significant) $t$ statistic. Hence, the researcher may be tempted to reject $H_0$ at nominal level $\alpha$ if $T_{max} > t_{1-\alpha/2} (n-k-1)$.}

The criterion $T_{max} > t_{1-\alpha/2} (n-k-1)$ amounts to maximizing evidence subject to snooping, which is a special case of manipulating evidence subject to snooping, MESSing in short after \citet{HasslerPohle2022}. This will distort the probability of a type I error. I quantify the expected distortion  experimentally. Under the null hypothesis of no causality I simulated $x_t$ and $y_t$ as two standard normal  iid sequences independent of each other. The sample size was $n=100$ with $k=2,3,4$. Tests were performed at level $\alpha = 0.05$ and $\alpha = 0.1$. The experiment was performed by means of the open source software R, the number of replications was $10^4$. In Table\,\ref{Tab_causal} I report the relative frequency of rejections. The size of the invalid max-test $\max \left\{|T_1 |, \ldots, |T_k | \right\}$ is all the more distorted the larger $k$ is. Such size distortions can  be avoided by employing the appropriate test designed for the overall $H_0$, which is of course the standard $F$ test. Let $F_k$ be the $F$ statistic testing for the joint null hypothesis of no Granger causality. The corresponding rejection frequencies building on the $F$ distribution with $k$ and $n-k-1$ degrees of freedom are given in the last line of Table\,\ref{Tab_causal}; they are reasonably close to the nominal $\alpha$.

\begin{table} [h!]

	\centering	
		\begin{threeparttable}     
		\caption{Relative frequency (in \%) of rejecting true $H_0$}
		\label{Tab_causal}
		\renewcommand{\arraystretch}{1.0}                                      
		\setlength{\tabcolsep}{5.5pt}
		\begin{tabular}{c|ccc|ccc}
			\vspace{0cm}
			& & $\alpha = 5 \%$ & & & $\alpha = 10 \%$ &  \\ 
			& $k=$2 & $k=$3 & $k=$4 & $k=$2 & $k=$3 & $k=$4\\ 
			\hline
			$T_{max}$ & 9.61 & 13.72 & 18.64 & 18.96 & 26.37 & 34.08 \\ 
			$F_k$ & 5.03 & 4.79 & 5.05 & 9.91 & 9.72 & 9.82 \\
		\end{tabular}

		\begin{tablenotes}													
			\small													
			\item \emph{Note} Rejection if $T_{max} > t_{1-\alpha/2} (n-k-1) $ for $n=100$ or $F_k > F_{1-\alpha}(k, n-k-1)$.
		\end{tablenotes}	
	\end{threeparttable}     
\end{table}

%

The lesson from this subsection is that preliminary significance snooping ruins subsequent significance tests. The larger the number of tests  that are screened for significance (i.\,e., the larger $k$ in our notation), the less reliable is the significance. In Remark~\ref{rem_4}, I present exact probability analogues to the frequencies of rejection of $T_{max}$. They are of the form $1 - (1 - \alpha)^k  = k \alpha - \frac{k (k-1)}{2} \alpha^2 - \cdots - (-\alpha)^k$ and very well  explain the experimental frequencies of rejection given in Table\,1. 	Generally for small $\alpha$, the dominating term  is $k \alpha$ resulting in growing size distortions with growing $k$.

\subsection{Post-model-selection inference}

The next example is inspired by \citet{LeebPoetscher2005}.

{\sc Example 3} \emph{Consider the regression model ($i=1,2, \ldots, n$)
	\[
	y_i = \delta + \beta x_i + \gamma z_i + \varepsilon_i \, , \quad \varepsilon_i  \sim \mathcal{N} (0, \sigma^2) \,,
	\]
 where the Gaussian errors are independent of the regressors. The exogeneous regressors $x_i$ and $z_i$ are correlated,
	\[
	\rho := \frac{Cov(x_i  ,z_i)}{\sqrt{Var(x_i) Var (z_i)}} \neq 0 \, ,
	\]
The null hypothesis of interest is	$H_0$: $\beta=\beta_0$. There are reasons to include the covariate $z_i$ (to ensure unbiasedness in case of $\gamma \neq 0$) as well as to exclude it  (to increase efficiency if $\gamma = 0$), resulting in unrestricted or restricted estimation, respectively,
 \begin{eqnarray*}
(U) \quad & & y_i = \widehat{\delta}_{U} + \widehat{\beta}_{U} x_i + \widehat{\gamma}_{U} z_i + \widehat{\varepsilon}_i^{U} \, , \\	(R) \quad & & y_i = \widehat{\delta}_{R} + \widehat{\beta}_{R} x_i + \widehat{\varepsilon}_i^{R}  \, .
\end{eqnarray*}
The ideal but infeasible estimator hence is
\[
\widetilde \beta := \left\{  \begin{array}{cc} \widehat{\beta}_{R} , & \mbox{if } \gamma = 0 \\ \widehat{\beta}_{U} , & \mbox{if } \gamma \neq 0  \end{array} \right. .
\]
 Since one does not know whether $\gamma = 0$ or not, one may be tempted to let the data speak and allow for a preliminary model selection step. The post-model-selection (PMS) estimator for $\beta$ becomes
\[
\widehat \beta_{PMS} := \left\{  \begin{array}{cc} \widehat{\beta}_{R} , & \mbox{if (R) is selected} \\ \widehat{\beta}_{U} , & \mbox{if (U) is selected}  \end{array} \right. .
\]	}

Under the classical assumptions maintained in Example~3, the $t$ statistic testing for $\beta = \beta_0$ follows a $t(n-2)$ distribution if the restricted model and $H_0$ are true, while $t(n-3)$ applies under model ($U$) and $H_0$. What is the effect of the model selection step on subsequent inference about $\beta = \beta_0$? The good news is that under consistent model selection it holds that
\[
\lim_{n \to \infty} \p (\widehat \beta_{PMS} = \widetilde \beta) = 1 \, ,
\]
where convergence is pointwise, not uniform; see \citet{Poetscher1991}. The bad news is that the limiting distribution of $\widehat \beta_{PMS}$ (and its $t$ statistic) depends on whether $\gamma =0 $ or not, notwithstanding consistent model selection. According to \citet{LeebPoetscher2005}, the PMS effect will be all the stronger the larger $|\rho|$ is, as long as the model ($U$) is ``close''  to ($R$), where ``closeness'' is relative to the sample size. 

\begin{figure}[h!]
	\noindent \centering{}\includegraphics[scale=0.5]{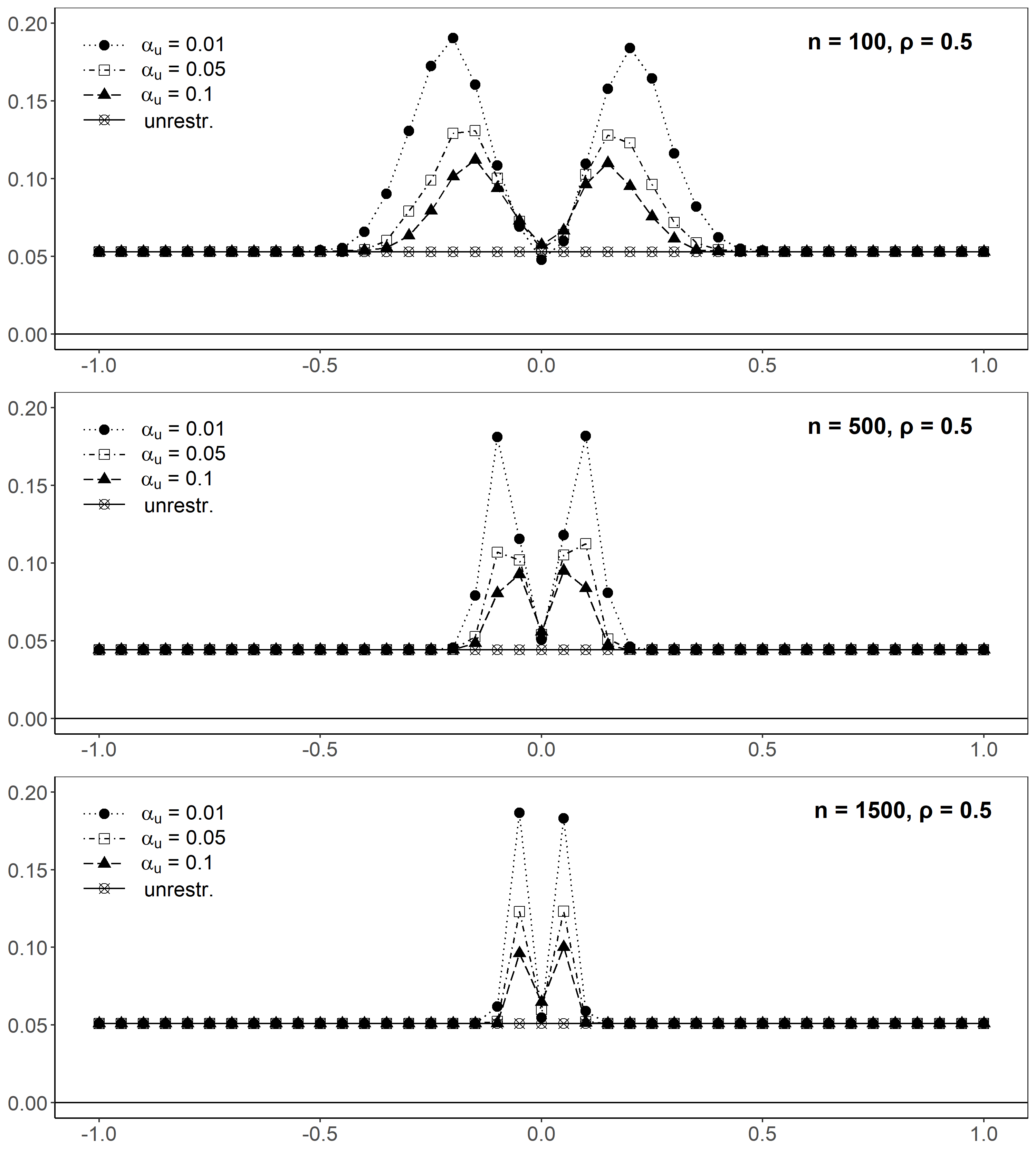}
	
	\vspace{-0cm}
	
	\caption{Rejection frequencies ($\alpha = 0.05$) of unrestricted estimation ($\widehat{\beta}_U$)  and after model selection ($\widehat{\beta}_{PMS}$) plotted against $\gamma$; $\rho = 0.5$}
	
\end{figure}

\begin{figure}[h!]
	\noindent \centering{}\includegraphics[scale=0.5]{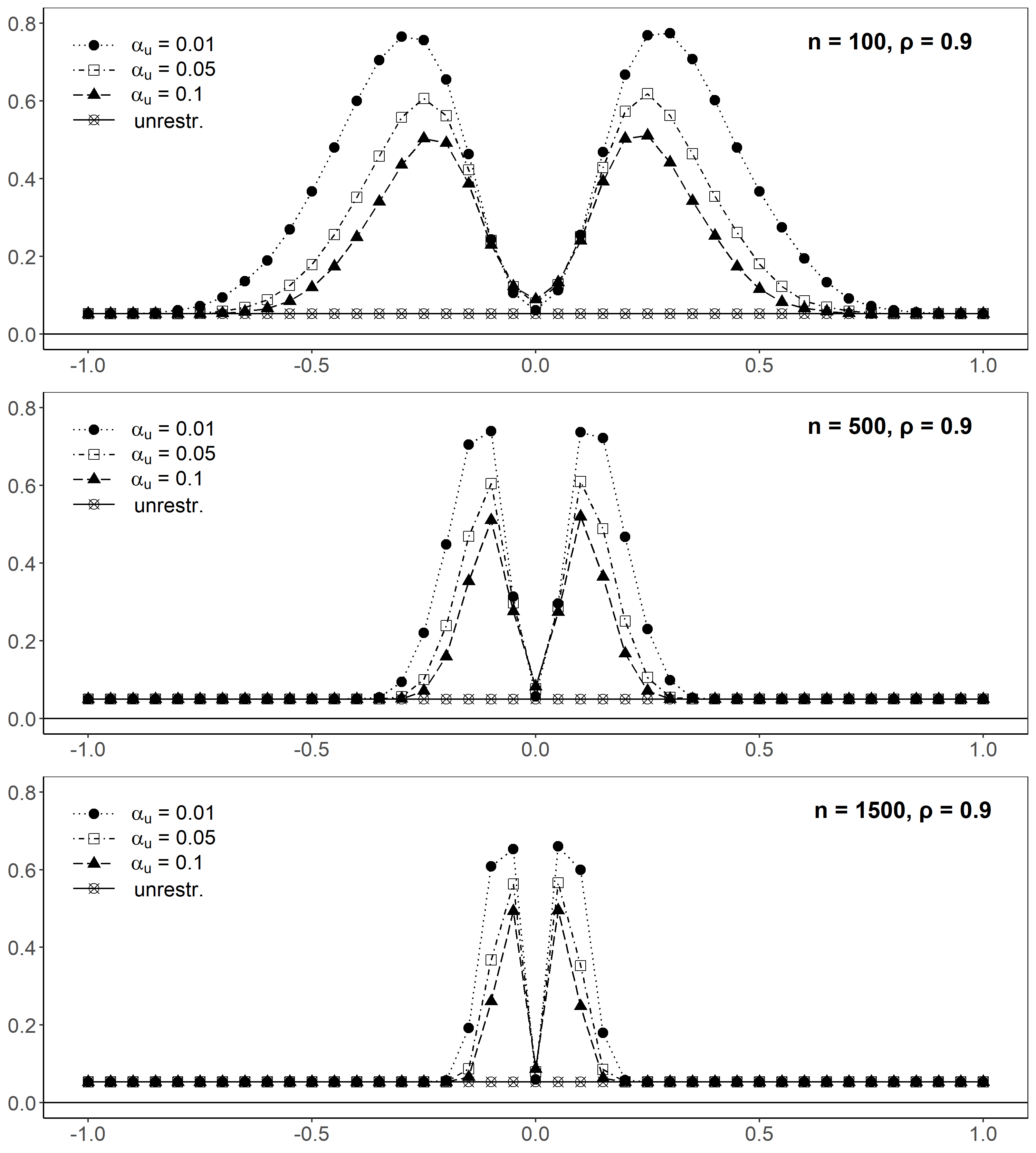}

	\caption{Rejection frequencies ($\alpha = 0.05$) of unrestricted estimation ($\widehat{\beta}_U$)  and after model selection ($\widehat{\beta}_{PMS}$) plotted against $\gamma$; $\rho = 0.9$} 
	
\end{figure}

To quantify the PMS effect on subsequent inference we simulate the model from Example~3 with 
\[
\left(\begin{array}{c} x_i \\ z_i \\ \varepsilon_i \end{array} \right) \sim \mathcal{N}_3 \left( \left(\begin{array}{c} 0 \\ 0 \\ 0\end{array}\right), \left(\begin{array}{ccc} 1 & \rho & 0 \\ \rho & 1& 0 \\ 0 & 0 & 0.5 \end{array} \right) \right) .
\]
The slope parameter of interest is $\beta=1$, and we test for this true null hypothesis at level $\alpha = 0.05$. To select the regression equation, a $t$ test for $\gamma=0$ is performed at level $\alpha_{U} \in \{0.01, 0.05, 0.10\}$:
\[
\widehat \beta_{PMS} = \left\{  \begin{array}{cc} \widehat{\beta}_{R} , & \mbox{if } |t_\gamma| \leq t_{1-\alpha_{U}/2} (n-3)\\ \widehat{\beta}_{U} , & \mbox{if } |t_\gamma| > t_{1-\alpha_{U}/2} (n-3)  \end{array} \right. .
\]
The experimental frequency of rejection from 5,000 replications for $\rho = 0.5$ is given for a grid of values of $\gamma$ in Figure\,1. We observe massive size distortions up to a rejection frequency of 20\,\% symmetrically around $\gamma =0$. The distortion is all the larger, the smaller $\alpha_{U}$ is. As a benchmark: Without model selection the inference relying on $\widehat{\beta}_U$ is valid with experimental rejections close to the nominal 5~\% level. Increasing the sample size from $n=100$ to $n=500$ or $n=1,500$  reduces the maxima of distortion only very little; they simply move closer to $\gamma =0$, where no distortion occurs.\footnote{Asymptotically, the size distortion vanishes for \emph{fixed} $\alpha_U$, while the model selection is of course not consistent for fixed $\alpha_U$.} The analogous picture arises for $\rho=0.9$ in Figure\,2, only that the distortions are even much more extreme. Valid PMS inference is a challenging problem, see for instance \citet{Berketal2013}.

Note that Examples\,2 and 3 are related. In textbooks, the teaching of model selection steps is well-intentioned and not motivated by SHARKing or MESSing. Still, it is difficult to rule out in practice that the final specification used for inference was determined such to produce highly significant $p$-values. As \citet{HasslerPohle2022} put it: ``there may in practice be a smooth transition from model selection to MESSing''.

\subsection{Discussion}

My reading (and writing) of Examples\,1 through 3 is related to a new field of epistemology called ``agnotology'' (from the Greek word ``agnosis'') that is devoted to ``the cultural production of ignorance (and its study)'', see \citet[p.~1]{Proctor2008}. \citet[p.~3]{Proctor2008} distinguished several kinds of ignorance, one of them being ``a deliberately engineered and \emph{strategic ploy} (or active construct)'', which ``can be made or unmade, and science can be complicit in either process.'' One of his ``favorite examples of \emph{agnogenesis} is the tobacco industry's efforts to manufacture doubt about the hazards of smoking'' (\citet[p. 11]{Proctor2008}, see also \citet{Proctor1995}). In addition to such actively created ignorance, the chapters in \citet{KouranyCarrier2020a} focus on passively constructed ignorance, one case being \emph{methodology-created ignorance}, see \citet[p. 16]{KouranyCarrier2020b}.

I consider Examples\,1 through 3 to illustrate the category of methodology-created ignorance: Even benevolent researchers may contribute to public ignorance by means of misleading significance tests in consequence of $p$-hacking, MESSing or PMS inference. Examples\,1 through 3 are characterized by less information or even growing confusion as a result of an increasing number of statistical tests. Clearly, the more hypotheses are tested with the same data or in the process of data snooping, the less convincing is statistical significance.  This reminds me of the recent pandemic. A quick internet search (carried out in May 2022) returns over 50 special issues on COVID-19 published between 2020 and 2022 in a variety of scientific journals (from fields like economics over social sciences and   psychology to statistics, not accounting for medicine, virology and related fields). This means that hundreds of empirical studies have been published building on COVID related data; many of them offer statistically significant results, and necessarily many of them analyse similar or identical data sets -- that did not arise from random sampling.

As indicated in Section\,2,  more fundamental objections against significance testing have been brought up, too, often by statisticians advocating a Bayesian approach. The problem of $p$ hacking, however,  is not automatically alleviated by Bayes techniques, see \citet{simonsohn2014}.  Further, Bayesian inference crucially hinges on prior probability assumptions that may be hard to justify in practice.

\section{Conclusions}

I believe that sheer fraud is rarely encountered in scientific studies and publications. Nevertheless, \citet{ioannidis2005} argued that  \emph{most published research findings are false}. Even if data are not manipulated and results are reported as grinded out by the computer (``facts''), the consequences may defective (``fake''), because the inferential tools are not applied in a proper manner. In Section\,3, I presented 3 examples how and when nonsensical or exaggerated significance may  occur in consequence of incorrect application.

Incorrect applications of statistical significance testing may have two roots. A) First, they may slip in unwantedly by ignorance of researchers who are otherwise in good faith; statistical pitfalls not only lurk in scientific publications but are even more likely to be encountered in everyday life studies in the fields of e.\,g., business, medicine, ecology and so on. B) Second, incorrect applications  may be  purposeful and exploited by otherwise smart researchers in order to generate  dazzling, highly significant results. The latter practice may be spurred by the so-called publication bias first addressed by \citet[p. 30]{Sterling1959}: ``[...]  research which yields nonsignificant results is not published. Such research being unknown to other investigators may be repeated independently until eventually by chance a significant result occurs - an `error of the first kind' - and is published.''; see also \citet{Sterlingetal1995}. Hence, there are  strong incentives for $p$ hacking, see \citet{Simonsohnetal2014}. This clearly calls for action. To encounter the first case A), better and more careful statistical training is required to create awareness when and how significance tests are valid. Further,  \citet[p.~703]{HirschauerGruenerMusshoffBecker2019} ``suggest twenty immediately actionable steps to reduce widespread
inferential errors'' related to significance testing; the first suggestion is ``Do not use $p$-values either if you have a non-random sample [...], 
$p$-values are not interpretable for non-random samples''. To account for the second case B),  it must be demanded that  ``Researchers should disclose the number of hypotheses
explored during the study, all data collection decisions,
all statistical analyses conducted, and all p-values computed'' (\citet[p. 132]{WassersteinLazar2018}). Since it is hard to control whether researchers  comply to such a policy, it is important to overcome the  publication bias; \citet{Sterlingetal1995}  suggested that empirical studies should be accepted for publication if they tackle a relevant or interesting research question with adequate methods and data, irrespective of the level of significance of the outcome; see also \citet{Simmonsetal2011}.

Examples~1 through 3 are not meant to be a plea against empirical research and statistical significance, but against unthinking repetition of research rituals that may not only be useless but even harmful, namely \emph{agnogenetic}.  Sometimes less may be  more, and in this note I tried to identify such cases.

\section*{Technical remarks}


\begin{remark} \label{rem_1} \emph{The value of the test statistic is
	\[
	z = \sqrt{100,000} \, \frac{0.08804 - 0.09}{\sqrt{0.09 \, (1- 0.09)}} = - 2.16578 \, .
	\]
Relying on the approximate normal distribution  the $p$-value of a  two-tailed test becomes $P=0.015 < 0.05$.}
	\end{remark}

\begin{remark} \label{rem_2} \emph{What is more, I performed a test for the null hypothesis that each number 1 through 100 has the equal probability of 0.01. The $\chi^2$ test statistic with 99 degrees of freedom rendered a $p$-value of $P=0.5229$, which does not suggest to reject.}  	\end{remark}


\begin{remark} \label{rem_4} \emph{To shed some light on the figures from the  computer experiment collected in Table\,1, I now construct a theoretical thought experiment. Consider the linear regression model in matrix form,
		\[
		\vy=\mX  \vbeta + \vu \, , \quad \vu \sim \mathcal{N}_n (\vzeros, \sigma^2 \mI_n) \, , 
		\]
		where $\sigma^2$ is known and $\mX$ is $n \times k$ with columns $\vx_{\bcdot h}$, $h=1, \ldots,k$. By orthogonality,  $\vx_{\bcdot h}^\prime \vx_{\bcdot g} = 0$ for $h \neq g$, one has
		\[
		\left( \mX^\prime \mX \right)^{-1} = \diag\left(\frac{1}{\vx_{\bcdot 1}^\prime \vx_{\bcdot 1}}, \ldots, \frac{1}{\vx_{\bcdot k}^\prime \vx_{\bcdot k}}\right).
		\]
		The $t$ type ratios become
		\[
		\tau_h := \frac{\widehat \beta_h - 0}{\sqrt{\frac{\sigma^2}{\vx_{\bcdot h}^\prime \vx_{\bcdot h}} }} =\frac{ {\widehat \beta_h} \sqrt{{\vx_{\bcdot h}^\prime \vx_{\bcdot h}}}}{\sigma} \ {\sim} \ \mathcal{N}(0,1),
		\]
		and $\tau_h$ and $\tau_g$ are independent. 	As in Example~2, I maintain the null hypotheses ($h=1, \ldots, k$) $H_0^{(h)}$ that $ \beta_h = 0$. 
		Under the null and the maintained assumptions a valid two-tailed test at significance level $\alpha$ builds on the rejection of $H_0^{(h)}$ if $|\tau_h | > z_{1-\alpha/2}$. As in Example~2, consider a researcher striving for an increase in power. Hence, he or she  does some prior data analysis in that he or she looks at all $k$ test statistics and chooses for testing the one that is largest in absolute value:
		\[
		\tau_{max} := \max \left\{|\tau_1 |, \ldots, |\tau_k | \right\} .
		\]
		The intersection $H_0$ is rejected if any $H_0^{(h)}$ is rejected. We assume a researcher unaware of the effect of MESSing: Since $\tau_{max} $ picks the most significant $|\tau_h |$, he or she rejects $H_0$ at level $\alpha$ if $\tau_{max} > z_{1-\alpha/2}$. For given $k$, one obtains as probability of a type I error when testing at nominal level $\alpha$ due to independence:
		\begin{eqnarray*}
			\p (\tau_{max} > z_{1-\alpha/2})	 &=& 1 - \p (\tau_{max} \leq z_{1-\alpha/2}) \\
			&=& 1 - \p (|\tau_1 | \leq z_{1-\alpha/2} \, ,  \ldots, |\tau_k | \leq z_{1-\alpha/2}) \\ &=& 1 -  \p (|\tau_1 | \leq z_{1-\alpha/2}) \cdots  \p ( |\tau_k | \leq z_{1-\alpha/2}) \\ &=& 1 - (1-\alpha)^k .
		\end{eqnarray*}
		Binomial expansion yields $(1-\alpha)^k = 1 + k (- \alpha)^1 + \frac{k(k-1)}{2} (- \alpha)^2 + \ldots + k (-\alpha)^{k-1} + (-\alpha)^{k}$.} 
	\begin{table} [h!]
		
		\centering	
		\begin{threeparttable}     
			\caption{Probability (in \%) of rejecting true null hypothesis}
			\renewcommand{\arraystretch}{1.0}                                      
			\setlength{\tabcolsep}{5.5pt}
			
			\begin{tabular}{c|ccc|ccc}
				& & $\alpha = 5 \%$ & & & $\alpha = 10 \%$ & \\ 
				& $k=$2 & $k=$3 & $k=$4 & $k=$2 & $k=$3 & $k=$4\\ 
				\hline
				$\tau_{max}$ &   9.75 & 14.26 & 18.55	& 19.00 & 27.10 & 34.39\\
				
			\end{tabular}
			\begin{tablenotes}													
				\small													
				\item \emph{Note} Rejection if  $\tau_{max} > z_{1-\alpha/2}  $.
			\end{tablenotes}	
		\end{threeparttable}     
	\end{table}	
	\end{remark}

\bibliography{literature_unlucky13}

\end{document}